\documentclass[aps,twocolumn, superscriptaddress , amsmath, amssymb, showkeys,reprint, nofootinbib]{revtex4-1}
\usepackage{graphicx}
\usepackage{dcolumn}
\usepackage{bm}
\usepackage{upgreek}
\usepackage{times}
\usepackage{mathrsfs}
\usepackage{multirow}
\usepackage{fancyhdr} 
\usepackage{hyperref} 
\usepackage[version=4]{mhchem} 

\begin{document}

\title{Structural dynamics in hybrid halide perovskites: Bulk Rashba splitting, spin texture, and carrier localization}
\author{Chao~Zheng}
\affiliation{Department of Materials Science and Engineering, McMaster University, 1280 Main Street West, Hamilton, Ontario L8S 4L8, Canada}
\author{Shidong Yu}
\affiliation{Department of Materials Science and Engineering, McMaster University, 1280 Main Street West, Hamilton, Ontario L8S 4L8, Canada}
\affiliation{College of Materials Science and Engineering, Jilin University, 2699 Qianjin Street, Changchun City, 130012, China}
\author{Oleg Rubel}
\email[]{rubelo@mcmaster.ca}
\affiliation{Department of Materials Science and Engineering, McMaster University, 1280 Main Street West, Hamilton, Ontario L8S 4L8, Canada}

\date{\today}

\begin{abstract}
The extended charge carrier lifetime in hybrid halide perovskites was attributed to a quasi-indirect band gap that arises due to a Rashba splitting in both conduction and valence band edges. In this paper, we present results for an effective relativistic band structure of \ce{(CH3NH3)PbI3} with the focus on the dispersion of electronic states near the band edges of \ce{(CH3NH3)PbI3} affected by thermal structural fluctuations. We establish a relationship between the magnitude of the Rashba splitting and a deviation of Pb-atom from its centrosymmetric site position in the \ce{PbI6} octahedron. For the splitting energy to reach the thermal energy $k_\text{B} T\approx26$~meV (room temperature), the displacement should be of the order of $0.3$~{\AA}, which is far above the static displacements of Pb-atoms in the tetragonal phase of \ce{(CH3NH3)PbI3}. The significant dynamic enhancement of the Rashba splitting observed at earlier simulation times (less than 2~ps) later weakens and becomes less than the thermal energy despite the average displacement of Pb-atoms remaining large (0.37~{\AA}). A randomization of Pb-displacement vectors and associated cancelation of the net effective magnetic field acting on electrons at the conduction band edge is responsible for reduction of the Rashba splitting. The lattice dynamics also leads to deterioration of a Bloch character for states in the valence band leading to the subsequent localization of holes, which affects the bipolar mobility of charge carriers in \ce{(CH3NH3)PbI3}. These results call into question the quasi-indirect band gap as a reason for the long carrier lifetime observed in \ce{(CH3NH3)PbI3} at room temperature. Analysis of spin projections and the spin overlap at the band edges also rules out the spin helicity as a possible cause for a long lifetime of optical excitations in perovskite structures. An alternative mechanism involves dynamic localization of holes and their reduced overlap with electrons in reciprocal space.
\end{abstract}


\maketitle

%
%
\section{Introduction}\label{Sec:Introduction}

Hybrid halide perovskites, with \ce{(CH3NH3)PbI3} being a prominent member, has attracted enormous interest as a solar cell absorber material \cite{Kojima_JACS_131_2009,Lee_S_338_2012,Gratzel_NM_13_2014,Bush_NE_2_2017,Singh_AEM_8_2018}.  One unique property of this class of materials is that they combine benefits of direct and indirect semiconductors featuring both a long carrier life time in excess of 100~ns \cite{Yamada_JACS_137_2015,Bi_SA_2_2016,Bi_JPCL_7_2016,Handa_JPCL_8_2017} and a sharp absorption edge \cite{Ziang_OME_5_2015}. These features are attributed to the presence of Rashba splitting, which occurs at the valence and conduction band edges of hybrid halide perovskites \cite{Even_PSSRRL_8_2014,Zheng_NL_15_2015,Kepenekian_AN_9_2015}.

\begin{figure}[t]
	\includegraphics{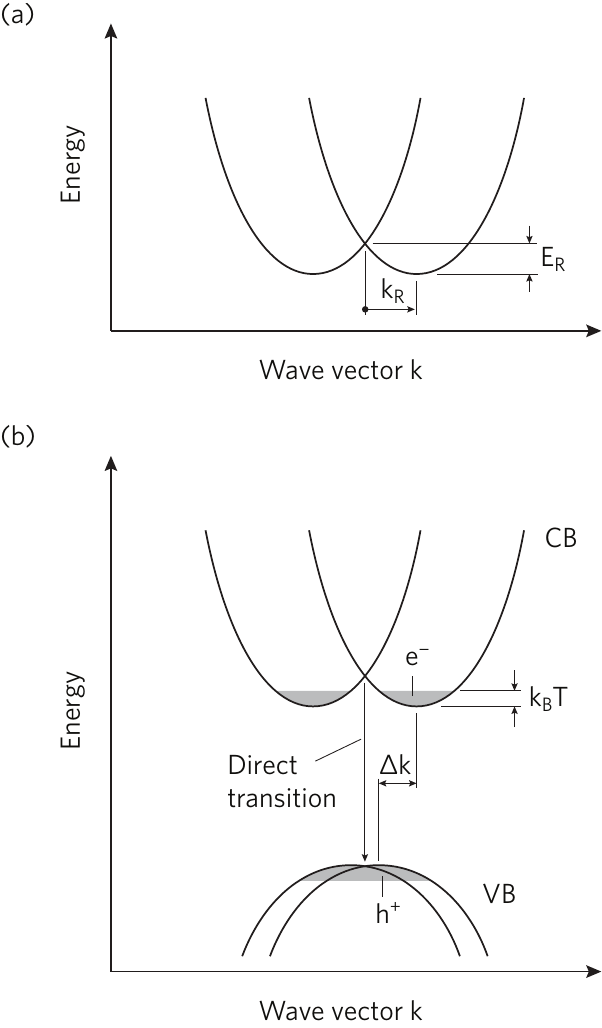}\\
	\caption{(a) Schematic Dresselhaus-Rashba splitting with $E_\text{R}$ showing the magnitude of splitting. (b) Although the band gap is indirect ($\Delta k$), the final temperature carrier statistic allows for direct transitions if $E_\text{R}<k_\text{B} T$.}\label{fig-1}
\end{figure}

The physics of the Dresselhaus-Rashba effect \cite{Dresselhaus_PR_100_1955,Rashba_FTTCP_2_1959} is linked to the spin-orbit coupling (SOC) due to an interaction between electron spin and an apparent magnetic field that arises from the electron moving in an electric field. Not all electronic states are directly susceptible to SOC. The coupling affects states that experience a net apparent magnetic field remaining after averaging over possible electron trajectories. In atoms, this leads to splitting of energy levels for states with the orbital angular momentum quantum number $\ell>0$. In solids, electron orbitals experience an additional crystal field.  If the crystal field lacks a central symmetry, Kramers' spin degeneracy is lifted leading to spin splitting and occurrence of an unusual band dispersion illustrated in Fig.~\ref{fig-1}a. The Dresselhaus-Rashba effect is actively studied in spintronics (spin currents and the spin Hall effect) and topological insulators \cite{Bihlmayer_NJP_17_2015}.

The extended carrier lifetime in hybrid halide perovskites was attributed to a quasi-indirect band gap that arises due to Rashba splitting in both conduction and valence band edges \cite{Katan_JMCA_3_2015,Azarhoosh_AM_4_2016,Wang_EES_10_2017} as shown in Fig.~\ref{fig-1}b. For this argument to stand, the magnitude of Rashba splitting must significantly exceed the solar cell operating temperature $(E_\text{R}>k_\text{B}T)$ to prevent the finite-temperature population statistics from enabling direct optical transitions (Fig.~\ref{fig-1}b). \citet{Niesner_PRL_117_2016} measured the valence band dispersion in \ce{(CH3NH3)PbI3} using an angle-resolved photoelectron spectroscopy and found a rather large magnitude of Rashba splitting of $E_\text{R}=0.16$ and 0.24~eV in orthorhombic (low temperature) and cubic (high temperature) phases, respectively, indicating that the splitting is enhanced by the lattice dynamics. Recent photogalvanic measurements  \cite{Niesner_PNASU_115_2018} provide more moderate values of the combined (valence and conduction band) Rashba splitting of $E_\text{R}\sim0.1$~eV that steadily increases with temperature.

Typically, the band structure of perovskites is calculated using a primitive unit cell (either pseudo-cubic, tetragonal, or orthorhombic structures), where atoms are relaxed to their lowest energy positions \cite{Menendez-Proupin_PRB_90_2014,Brivio_PRB_89_2014}. However, experimental studies of hybrid halide perovskites \cite{Whitfield_SR_6_2016,Bernard_JPCA_122_2018} clearly indicate a substantial dynamic disorder of atomic positions at room temperature, which should influence the electronic structure of these materials. For the aforementioned mechanism to operate, the lattice dynamics should produce a steady enhancement of the Rashba splitting and, \textit{at the same time}, preserve the Bloch character (wave vector $k$) for electronic states at the band edges. Theoretical studies of dynamic effects on the Rashba splitting started only recently \cite{Etienne_JPCL_7_2016,Monserrat_PRM_1_2017,Monserrat_arXiv_1711.06274,Etienne_JPCC_122_2018,McKechnie_PRB_98_2018}, and details on how the dispersion of electronic states near the band edges of \ce{(CH3NH3)PbI3} is affected by thermal structural fluctuations are still missing.

In this paper, we calculate an effective relativistic band structure of \ce{(CH3NH3)PbI3} taking the thermal disorder of atomic positions into account. The disorder is explicitly modelled via \textit{ab initio} molecular dynamics (MD) simulation performed for a large supercell. The supercell band structure is later unfolded to a primitive (pseudo-cubic) Brillouin zone that allows its direct comparison with experimental data. In addition, we explore a possible spatial localization of electronic states at the band edges caused by dynamical structural fluctuations. Robustness of electronic states at the band edges is an important functional requirement for photovoltaic applications.

%
%
\section{Computational details}\label{Sec:Method}

The Vienna \textit{ab-initio} simulation program (VASP) \cite{Kresse_PRB_54_1996,Kresse_CMS_6_1996} density functional theory \cite{Kohn_PR_140_1965} (DFT) package was employed in this work. A \citet*{Perdew_PRL_77_1996} (PBE) gradient approximation for the exchange-correlation functional was used in combination with the \citet{Grimme_JCP_132_2010} (D3) correction to capture long-range van der Waals interactions.

The original structure of tetragonal \ce{(CH3NH3)PbI3} was taken from \citet{Stoumpos_IC_52_2013} (database code ICSD 250739) with the following structural parameters: space group 108 (I4cm), $a=b=8.849$~{\AA}, $c=12.642$~{\AA}, fractional coordinates $u_\text{Pb}=(0,0,0)$, $u_\text{I1}=(0,0,0.24720)$, $u_\text{I2}=(0.21417,0.71417,0.00460)$. Once the \ce{PbI3} cage was set, methylammonium cations were added, and the structure was relaxed while maintaining symmetry of the \ce{PbI3} cage. Calculations were carried out using a $3\times3\times2$ \citet*{Monkhorst_PRB_13_1976} $k$-mesh for the primitive Brillouin zone of a tetragonal phase. The structural relaxation was performed by minimizing Hellmann-Feynman forces and stresses below 20~meV/{\AA} and 0.5~kbar, respectively. The cutoff energy for the plane-wave expansion was set at 400~eV. Relativistic effects (SOC) were omitted from the structure optimization but included later in band structure calculations. The resultant optimized structure had lattice parameters of $a=b=8.661$~{\AA} and $c=12.766$~{\AA} and can be accessed at the Cambridge crystallographic data centre (CCDC) under deposition no.~1870783.

\begin{figure}[t]
	\includegraphics{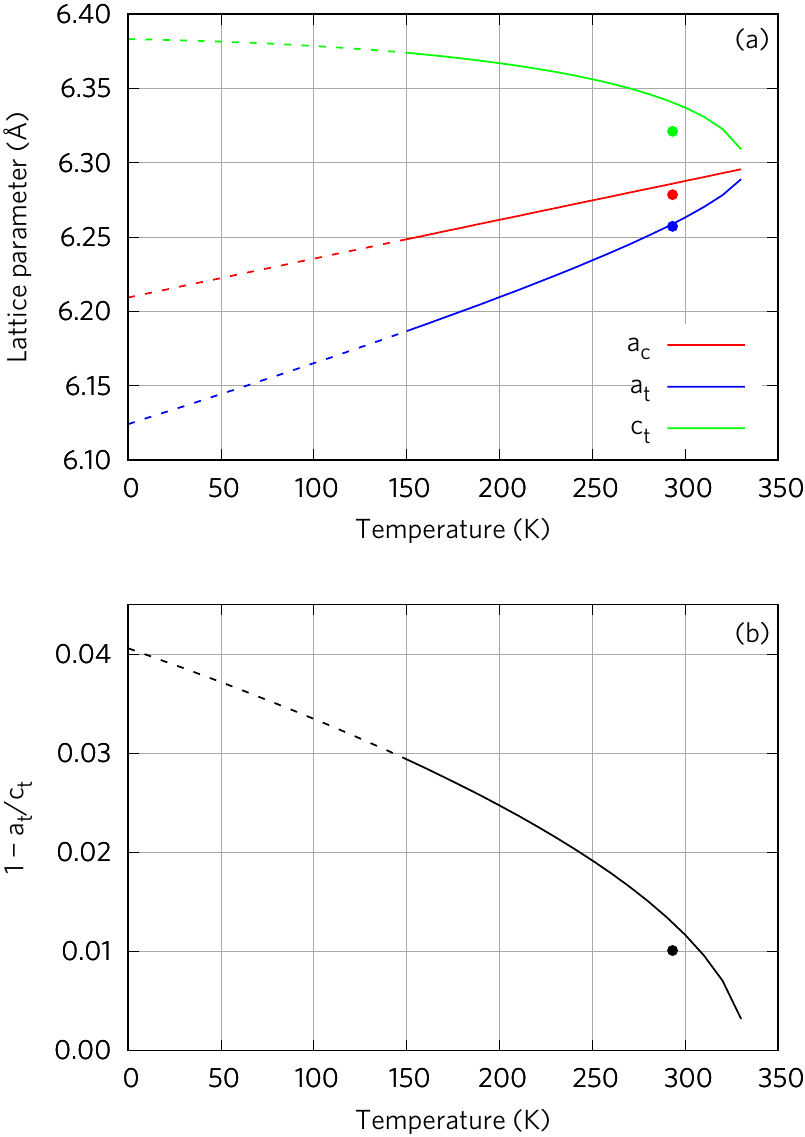}\\
	\caption{Finite temperature scaling of \ce{(CH3NH3)PbI3} (a) lattice parameters and (b) tetragonality starting with DFT-PBE+D3 results for 0~K. Data are shown for $T<150$~K even though the tetragonal structure is unstable in this temperature range. Experimental data \cite{Stoumpos_IC_52_2013} are shown with points.}\label{fig-2}
\end{figure}

Next we describe a semi-empirical scaling of the lattice parameters to achieve a finite-temperature structure of \ce{(CH3NH3)PbI3} that is self-consistent with PBE+D3 functional. The volume of the calculated tetragonal unit cell at 0~K yields a pseudo-cubic lattice parameter of $a_c(0~\text{K})=6.209$~{\AA}. Changes in the pseudo-cubic lattice parameter with temperature can be expressed as:
\begin{equation}\label{Eq:1}
 a_c(T) = a_c(0~\text{K})(1+\alpha_a T)
\end{equation}
with the linear expansion coefficient of $\alpha_a=4.21\cdot10^{-5}$~K$^{-1}$ inferred from the experimental data \cite{Whitfield_SR_6_2016} in the range of temperatures $150-350$~K (see Fig.~6c therein). Evolution of the pseudo-cubic lattice parameter with temperature is shown in Fig.~\ref{fig-2}a.

Tetragonality of the structure can be expressed as:
\begin{equation}\label{Eq:2}
 t = 1-a_t/c_t
\end{equation}
where $a_t=b_t$ and $c_t$ are tetragonally distorted lattice parameters that are related to conventional lattice parameters of the tetragonal structure via $a=a_t\sqrt{2}$ and $c=2c_t$. At 0~K the tetragonality is $t(0~\text{K})=0.04058$ as obtained from our DFT calculations. The tetragonality is significantly reduced with increasing temperature and amounts to $t(293~\text{K})=0.01$ \cite{Stoumpos_IC_52_2013}. The temperature-dependent tetragonality can be captured by a scaling function \cite{Whitfield_SR_6_2016}:
\begin{equation}\label{Eq:3}
 t(T) = t(0)
 \left(
 1-T/T_c
 \right)^{2\beta}
\end{equation}
with critical parameters $T_c=333$~K and $\beta=0.27$ taken from \cite{Whitfield_SR_6_2016} (see Fig.~8b therein). The temperature-dependent tetragonality parameter is shown in Fig.~\ref{fig-2}b. The temperature-dependent tetragonally distorted lattice parameters $a_t(T)$ and $c_t(T)$ (Fig.~\ref{fig-2}a) were obtained by combining the $t(T)$ and $a_c(T)$ dependencies assuming an equal volume for both cubic and tetragonal structures ($a_c^3=c_ta_t^2$) at a given temperature.

This approach yields finite-temperature tetragonal lattice parameters of $a=b=8.858$~{\AA} and $c=12.674$~{\AA} at 300~K that are self-consistent with PBE+D3. Atomic positions of this structure were further relaxed while maintaining symmetry of the \ce{PbI3} cage. The corresponding structure file can be accessed at CCDC under deposition no.~1870784.

The MD simulation was performed in two stages: first pre-heating and then a ``production" run. The initial structure was a $4\times4\times4$ supercell (Fig.~\ref{fig-10}a) with lattice parameters scaled to 300~K as described above. The symmetry (except for the translational one) was turned off during the simulation (VASP tag $\text{ISYM}=0$). Pre-heating from 0 to 300~K was performed in 500 steps (step size of 1~fs) using a linear ramp-up function (VASP tag $\text{SMASS}=-1$). Velocities were scaled every 10 MD steps. Accuracy of computed  Hellmann-Feynman forces was determined by the energy convergence criterion of $10^{-7}$~eV. Only one $k$-point at $\Gamma$ was used to sample the Brillouin zone. Atomic positions and velocities at the end of the preheating stage were taken as input for the production run. The production run took 2,200 steps (step size of 1~fs) of a constant energy MD (VASP tag $\text{SMASS}=-3$). Atomic positions during MD were stored every 10 steps.

Band structure calculations were performed taking SOC into account. Although the band gap is underestimated at the DFT-PBE level of theory, the band dispersion and the Rashba splitting should be properly captured. The band structure of supercells was unfolded to a primitive Brillouin zone corresponding to a pseudo-cubic structure. The unfolding was performed with a \texttt{fold2Bloch} utility \cite{Rubel_PRB_90_2014}.

A spin texture was analyzed in the following way. The spinor wavefunction is represented as a linear combination of spin up and down components:
\begin{equation}\label{Eq:4}
  | \psi \rangle = \alpha | \uparrow \rangle + \beta | \downarrow \rangle =
  \begin{pmatrix}
  \alpha\\
  \beta
  \end{pmatrix},
\end{equation}
that fulfill the normalization requirement $\alpha^2+\beta^2=1$. Spin projections for individual eigenstates
\begin{equation}\label{Eq:5}
  \langle S_{m} \rangle = \langle \psi | \sigma_m | \psi \rangle \quad (m=x,y,z)
\end{equation}
were computed along cartesian coordinates (VASP tag $\text{LORBIT}=11$, PROCAR file). Here $\sigma_m$ are Pauli matrices. The spin projections are related to spinor components as
\begin{subequations}
  \label{Eq:6-all} 
  \begin{eqnarray}
    \langle S_x \rangle & = & \alpha^*\beta + \beta^*\alpha,
    \\
    \langle S_y \rangle & = & i(\beta^*\alpha - \alpha^*\beta),
    \\
    \langle S_z \rangle & = & \alpha^2 - \beta^2.
  \end{eqnarray}
\end{subequations}
It is possible to determine $\alpha$ and $\beta$ from Eqs.~(\ref{Eq:6-all}) with the uncertainty of a phase factor $e^{i\theta}$, which neither affects the relative contribution of spin up/down components nor the spin overlap between two states. The uncertainty was resolved by constraining $\text{Im}(\alpha)=0$ that is consistent with eigenvectors of Pauli matrices.

The inverse participation ratio (IPR) $\chi$ was used as a measure of localization. It was evaluated on the basis of probabilities $\rho_n (E_i)$ of finding an electron with an eigenenergy $E_i$ within a muffin tin sphere centred at an atomic site $n$ \cite{Murphy_PRB_83_2011,Pashartis_PRA_7_2017}:
\begin{equation}\label{Eq:7}
   \chi (E_i) = \dfrac
   {\sum_{n} \rho_n^2 (E_i)}
   {\left[\sum_{n} \rho_n (E_i) \right]^2}~.
\end{equation}
Here the summation index $n$ runs over all atomic sites.

%
%
\section{Results and discussion}\label{Sec:Results}
\subsection{Static structures}

\begin{figure}[t]
	\includegraphics{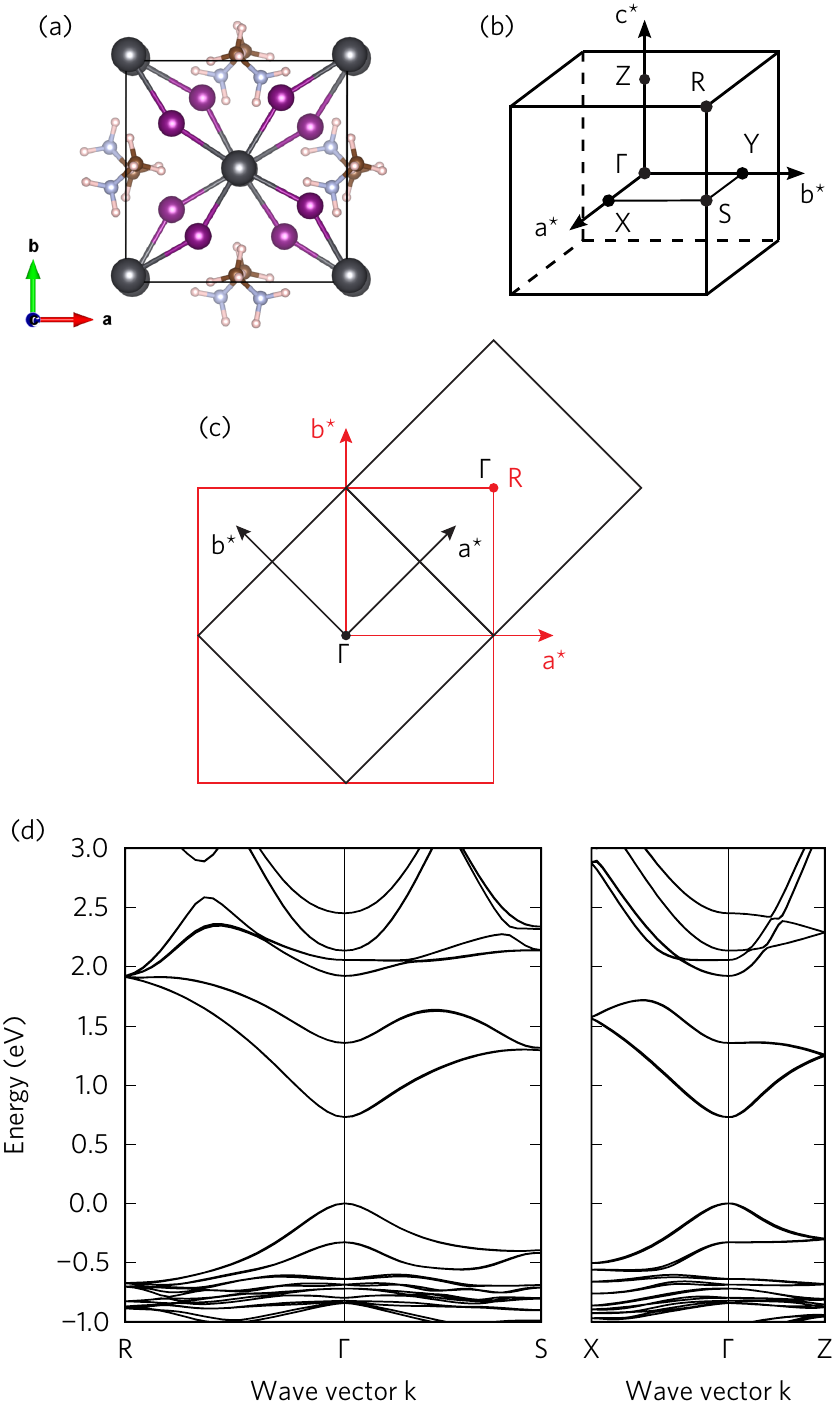}\\
	\caption{(a) Tetragonal unit cell, (b) Brillouin zone, (c) folding of the R-point into $\Gamma$ when the structure is transformed from pseudo-cubic (red) to tetragonal (black) cell, and (d) relativistic band structure of tetragonal \ce{(CH3NH3)PbI3} with lattice parameters scaled to 300~K and the symmetrized \ce{PbI3} cage. No sizable Rashba splitting is observed although the structure is non-centrosymmetric. The origin of the energy scale is set at the Fermi energy.}\label{fig-3}
\end{figure}

We begin with presenting the electronic structure of tetragonal \ce{(CH3NH3)PbI3} with lattice parameters scaled to $T=300$~K and the \ce{PbI3} cage symmetry (Fig.~\ref{fig-3}a) obtained from x-ray diffraction studies \cite{Stoumpos_IC_52_2013}. The quasi-direct band gap is typically observed at the R-point of the Brillouin zone (Fig.~\ref{fig-3}b). However, the R-point is folded into $\Gamma$ in the tetragonal phase (Fig.~\ref{fig-3}c). The corresponding band structure is presented in Fig.~\ref{fig-3}d and shows no Rashba splitting as also noticed in \cite{Etienne_JPCL_7_2016}. A small non-degeneracy occurs at the band edges due to the field from MA cations.

Experimentally, tetragonal \ce{(CH3NH3)PbI3} has a centrosymmetric Pb-site symmetry even though the whole structure lacks an inversion center. \citet{Etienne_JPCL_7_2016}, and \citet{Kepenekian_JPCL_8_2017} noted that Rashba splitting occurs due to breaking a \textit{site} inversion asymmetry. This symmetry argument explains the absence of the Rashba splitting in Fig.~\ref{fig-3}d.

\begin{figure}[t]
	\includegraphics{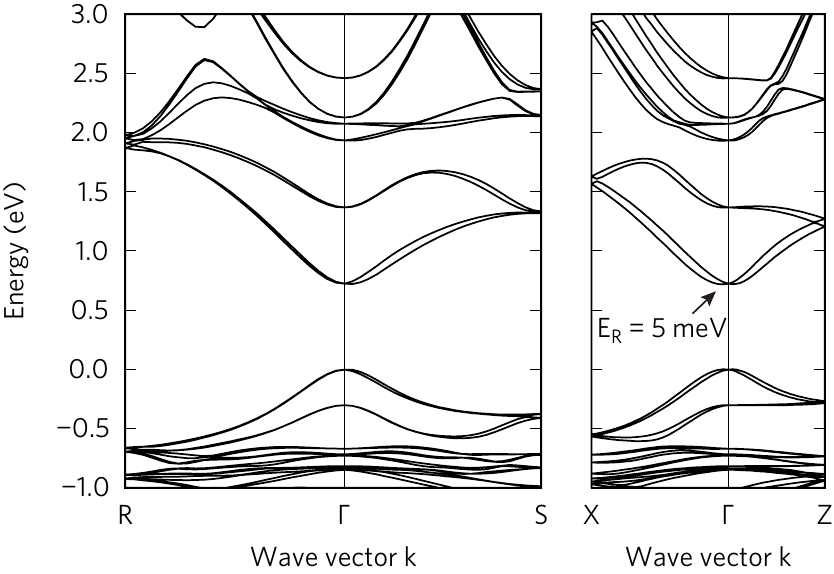}\\
	\caption{Relativistic band structure of tetragonal \ce{(CH3NH3)PbI3} with lattice parameters scaled to 300~K and relaxed \ce{PbI3} cage. The Rashba splitting is presents, but its magnitude is less that $k_\text{B} T$ at room temperature.}\label{fig-4}
\end{figure}

Next, we fully relax atomic positions in the tetragonal structure (CCDC deposition no.~1870791) and repeat the calculation of the band structure. The relaxed structure shows a signature of Rashba splitting at the $\Gamma$-point at the conduction band edge (CBE), although its magnitude is rather weak ($E_\text{R}=5$~meV, Fig.~\ref{fig-4}), which is comparable to $E_\text{R}=10$~meV found in relativistic quasiparticle calculations \cite{McKechnie_PRB_98_2018}. This result implies that the symmetry of the \ce{PbI3}-cage is broken after relaxation. The Rashba splitting is more prominent in the conduction band that is mostly comprised of Pb-$p$ orbitals (83\%) with a minor contribution from I-$p,s$ (14\%). The orbital character at the valence band edge (VBE) has significantly less contribution from lead: I-$p$ (70\%) and Pb-$s$ (30\%).

\begin{figure}[t]
	\includegraphics{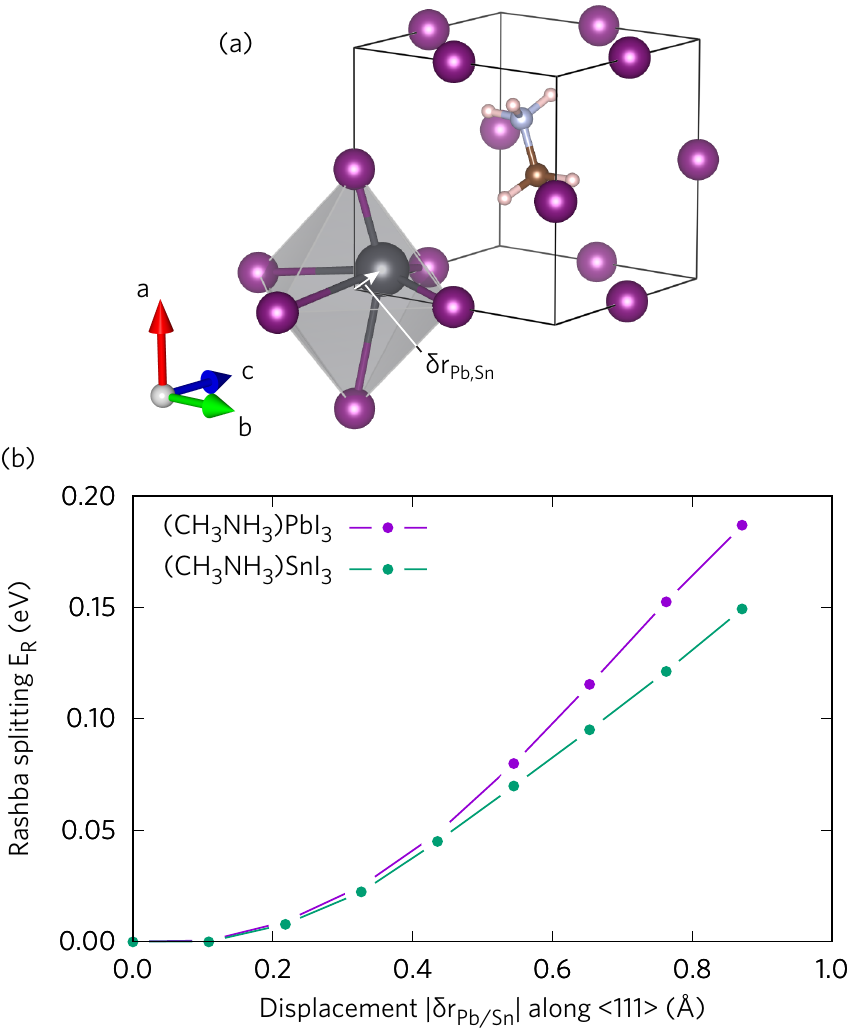}\\
	\caption{(a) Displacement vector $\delta \bm{r}$ of Pb or Sn-atom from the center of octahedra formed by iodine atoms. (b) Rashba splitting in a cubic structure as a function of the displacement of Pb or Sn-atom from its centrosymmetric position.}\label{fig-5}
\end{figure}

To quantify a displacement of the Pb-atom from the center of mass of a tetrahedron formed by iodine atoms (see Fig.~\ref{fig-5}a) we use the following expression
\begin{equation}\label{Eq:8}
 \delta \bm{r}_\text{Pb} = \bm{r}_\text{Pb} - \frac{1}{6} \sum_{i=1}^6 \bm{r}_{\text{I}_i}.
\end{equation}
Here $\bm{r}_\text{Pb}$ and $\bm{r}_{\text{I}_i}$ refer to the cartesian coordinates of the corresponding atomic species. The tetragonal structure with relaxed atomic positions has a displacement of about $\delta r_\text{Pb}\sim0.1$~{\AA}. Our results are consistent with the Pb-displacement value of 0.09~{\AA} reported by \citet{Liu_JPCL_7_2016} for the tetragonal phase. The displacement in a cubic structure is greater (about 0.25~{\AA} \cite{Yan_JPCC_120_2016}), which results in a larger magnitude of $E_\text{R}$ typically found in those structures \cite{Hu_JPCC_121_2017}.

To link the magnitude of the Rashba splitting $E_\text{R}$ with the displacement of Pb-atom, we performed calculations for a cubic structure ($a=6.29$~{\AA}) by varying $\delta \bm{r}_\text{Pb}$ with I-atoms fixed at their perfect positions expressed by the fractional coordinates $(1/2,0,0)$ and permutations of that. The most effective displacement direction for enhanced $E_\text{R}$ is along the diagonal of the cube $\langle 111 \rangle$ (Fig.~\ref{fig-5}a, CCDC deposition no.~1870792). The largest Rashba splitting is observed in the $k$-plane that is perpendicular to the displacement vector $\delta \bm{r}_\text{Pb}$; the splitting vanishes for $k$-paths oriented parallel to the displacement vector.

Results for $E_\text{R}(\delta r_\text{Pb})$ are shown in Fig.~\ref{fig-5}b. It is interesting that the same structure with Sn instead of Pb (used as a lead-free alternative to \ce{(CH3NH3)PbI3} perovskites \cite{Kamat_AEL_2_2017}) shows a comparable $E_\text{R}$ in spite of a drastic difference in the SOC constant between Pb and Sn ($\lambda_\text{SO}=0.91$~eV vs 0.27~eV \cite{Wittel_TCA_33_1974}, respectively). At first this result may look counterintuitive since it is commonly accepted \cite{Picozzi_FP_2_2014} that the Rashba parameter $\alpha_R$ is proportional to $\lambda_\text{SO}$, which leads to the much stronger dependence $E_\text{R}\propto \lambda_\text{SO}^2$. However, the SOC constant $\lambda_\text{SO}$ is determined for isolated atoms and serves as an approximation for orbitals that are significantly altered by chemical bonding. Comparable values of $E_\text{R}$ for Pb-based and Sn-based structures suggest that the vector product of an effective asymmetric electric field $\bm{E}_0$ experienced by electrons and their velocity $\bm{v}$ averaged over possible trajectories for a given orbital is similar in both cases.

For the splitting energy to reach $k_\text{B} T\approx26$~meV at room temperature, the displacement should be of the order $0.3$~{\AA} (Fig.~\ref{fig-5}b), which is far above the static displacements of Pb-atoms in the tetragonal phase of \ce{(CH3NH3)PbI3}. Does the lattice dynamics provide sufficient distortions for the Rashba splitting to exceed 26~meV?

\subsection{Dynamic structures}

\begin{figure}[t]
	\includegraphics{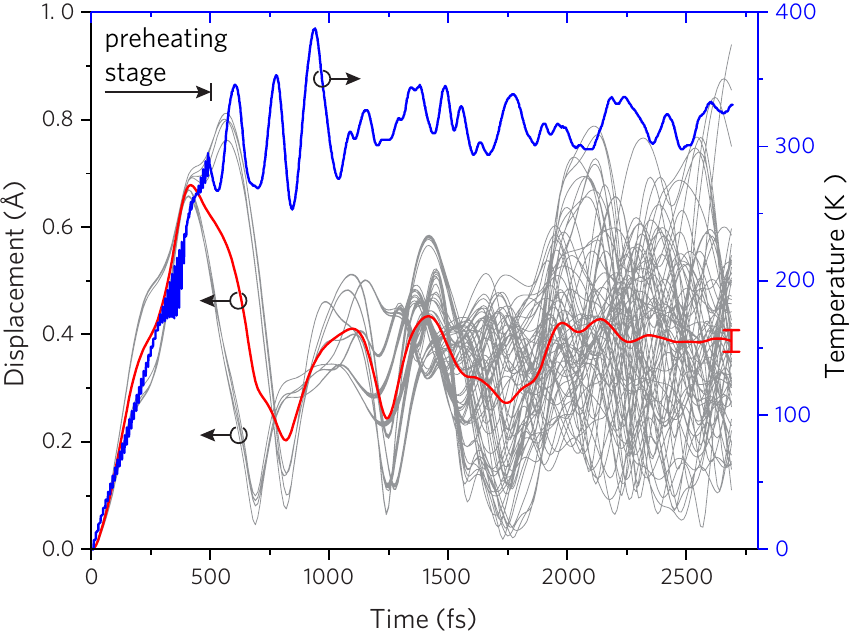}\\
	\caption{Time evolution of a displacement $\delta r_{\text{Pb}_i}$ of Pb-atoms from the center of their individual octahedra (grey curves, left axis), the average $\langle \delta r_\text{Pb} \rangle$ for all Pb-atoms in the supercell (red curve, left axis), and temperature $T$ of the MD ensemble (blue curve, right axis). First 500~fs correspond to the pre-heating stage. Displacements fully dephase after approximately 2000~fs and $\langle \delta r_\text{Pb}(t) \rangle$ stabilizes.}\label{fig-6}
\end{figure}

Now we turn our attention to a dynamic structure of \ce{(CH3NH3)PbI3}. The supercell size selected in our calculations greatly exceeds previous simulations \cite{Etienne_JPCL_7_2016,McKechnie_PRB_98_2018}, which provides a more realistic model for structural fluctuations and allows access to phonon modes that are otherwise not present in smaller cells.  Figure~\ref{fig-6} shows a time evolution of $\delta r_\text{Pb}$. The dynamic amplitude of $\delta r_\text{Pb}$ near to room temperature can reach 0.8~{\AA}, and the average displacement stabilizes near 0.37~{\AA}. According to Fig.~\ref{fig-5}, we would expect a much stronger Rashba splitting in those structures.

\begin{figure*}[t]
	\includegraphics{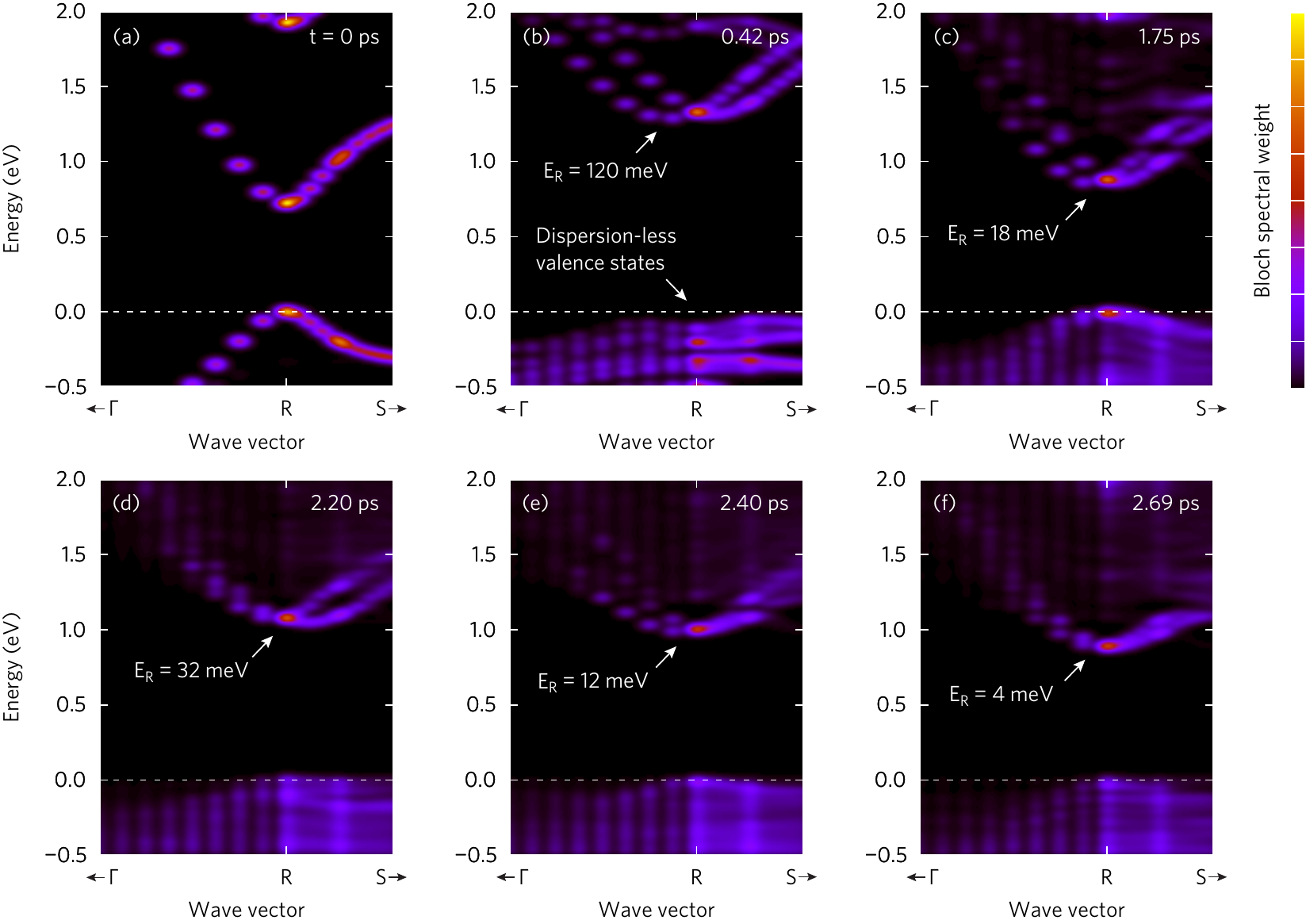}\\
	\caption{Effective band structure of $4\times4\times4$ tetragonal \ce{(CH3NH3)PbI3} supercell at different MD time snapshots:  (a) $t=0$, (b) 0.42~ps, (c) 1.75~ps, (d) 2.20~ps, (e) 2.40~ps, and (f) 2.69~ps. The Bloch character degrades at the VBE. Values for the Rashaba splitting $E_\text{R}$ were derived from all eigenvalues in the calculation, including those that do not appear on the $k$-path selected for the band structure plot. The strongest dynamical enhancement of $E_\text{R}$ is observed at the initial time steps. Data points were smeared using a Gaussian function with the standard deviation of $\sigma_k=0.0025~\text{\AA}^{-1}$ in $k$-space and $\sigma_E=25$~meV on the energy scale.}\label{fig-7}
\end{figure*}

Interpretation of band structure calculations of supercells is not straightforward due to a band folding. Here, we used an unfolding technique implemented in \texttt{fold2Bloch} utility \cite{Rubel_PRB_90_2014}. Effective band structures of the \ce{(CH3NH3)PbI3} supercell, taken at different MD simulation times, are shown in Fig.~\ref{fig-7} (structure files can be accessed at CCDC deposition no.~1870785$\!-\!$1870790). As anticipated, the initial ($t=0$) structure features no Rashba splitting (compare Fig.~\ref{fig-7}a and Fig.~\ref{fig-3}d) and serves as a benchmark for comparison with other structures. The Bloch character of primitive wave vectors $k$ (pseudo-cubic in this case) is well defined for the initial structure, which is indicative of periodicity of the wave function preserved throughout the supercell.

It should be noted that values of the Rashba splitting $E_\text{R}$ in Fig.~\ref{fig-7} are evaluated with respect to the CBE in the \textit{full} Brillouin zone
\begin{equation}\label{Eq:9}
 E_\text{R}=E_\text{CB}(R)- \min\left[E_\text{CB}(\bm{k})\right].
\end{equation}
The energy eigenvalue $\min\left[E_\text{CB}(\bm{k})\right]$ may not necessarily lie on the $\Gamma\!-\!R\!-\!S$ path selected for the band structures in Fig.~\ref{fig-7}, which can give an impression of an inconsistency between the band structure and $E_\text{R}$ values. At $t=0.42$~ps, for instance, the CBE unfolds into $\bm{k}=(1/2,-0.375,1/2)$ with the Bloch spectral weights of 62\%. Since this $k$-point does not belong to the $\Gamma\!-\!R\!-\!S$ path, the band structure in Fig.~\ref{fig-7}a does capture the full extent of the Rashba splitting, whereas $E_\text{R}$ does.

In dynamic structures we observe breaking of Kramers' degeneracy with the most pronounced Rashba splitting of $E_\text{R}\approx120$~meV (Fig.~\ref{fig-7}b) observed in the conduction band near the end of the pre-heating stage ($t=0.42$~ps). It is a manifestation of the dynamic Rashba splitting present even in centrosymmetric structures~\cite{Monserrat_arXiv_1711.06274,Niesner_PNASU_115_2018,McKechnie_PRB_98_2018}. This timeframe corresponds to the maximum average displacement in Fig.~\ref{fig-6} and sets an upper limit for dynamic $E_\text{R}$. It is instructive to analyze this result, but it can hardly be physically plausible since the MD displacements of Pb-atoms are still coherent at this time. The Rashba splitting gradually decreases as the dynamics of Pb-atoms becomes less coherent (Fig.~\ref{fig-7}d-f) even though the average displacement vector $\langle \delta r_\text{Pb}(t) \rangle$ remains steady at the MD time $t>2$~ps (Fig.~\ref{fig-6}).  This behavior suggests that not only the displacement magnitude but also a mutual orientation of displacement vectors play a role in determining $E_\text{R}$.

\begin{figure}[t]
	\includegraphics{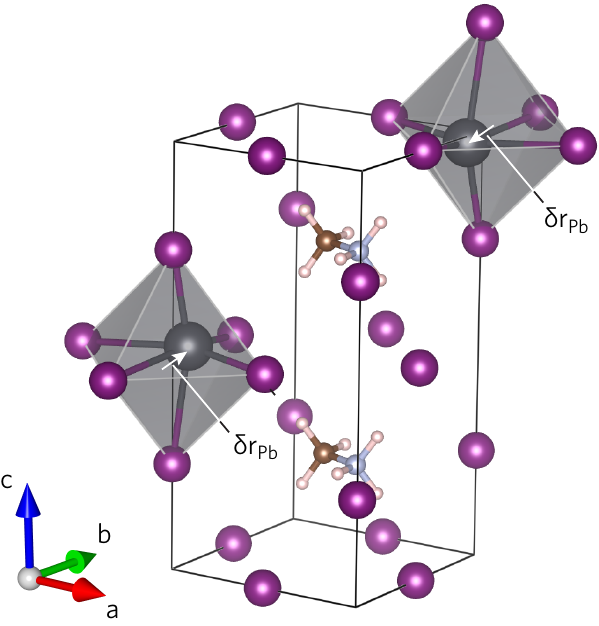}\\
	\caption{Perovskite structure with two Pb-atoms displaced in opposite direction by the same magnitude.}\label{fig-8}
\end{figure}

To verify this hypothesis, we created a supercell where Pb-atoms in two \ce{PbI6} octahedra had equal displacements by magnitude but opposite in direction ($\delta \bm{r}_\text{Pb1}=-\delta \bm{r}_\text{Pb2}$) as shown in Fig.~\ref{fig-8} (CCDC deposition no.~1870793). In spite of sizable displacements ($\delta r_\text{Pb}=0.5$~{\AA}), the Rashba splitting completely vanishes in this structure. To rationalize this result, we recall the effective asymmetric electric field $\bm{E}_0$ that causes Rashba splitting is aligned with the displacement $\delta \bm{r}_\text{Pb}$. When atoms are displaced in opposite directions, the electric field on each atomic site opposes each other leading to cancelation of the net effective electric field. This reasoning explains why dynamic structures have a small $E_\text{R}$ despite a relatively large average displacement.

Next, we discuss the valence band. In Fig.~\ref{fig-7}, the valence band becomes too disordered to make any conclusions about Rashba splitting there. This result suggests that $k$ is not a good quantum number for holes in the dynamic structure of \ce{(CH3NH3)PbI3}. The loss of Bloch character at the VBE undermines the argument about an indirect band gap (Fig.~\ref{fig-1}b), which implies that $\Delta k$ is well resolved. Instead, the prolonged carrier lifetime can originate from a weak overlap in real and reciprocal space between extended electron states and localized hole states.

\begin{figure*}[t]
	\includegraphics{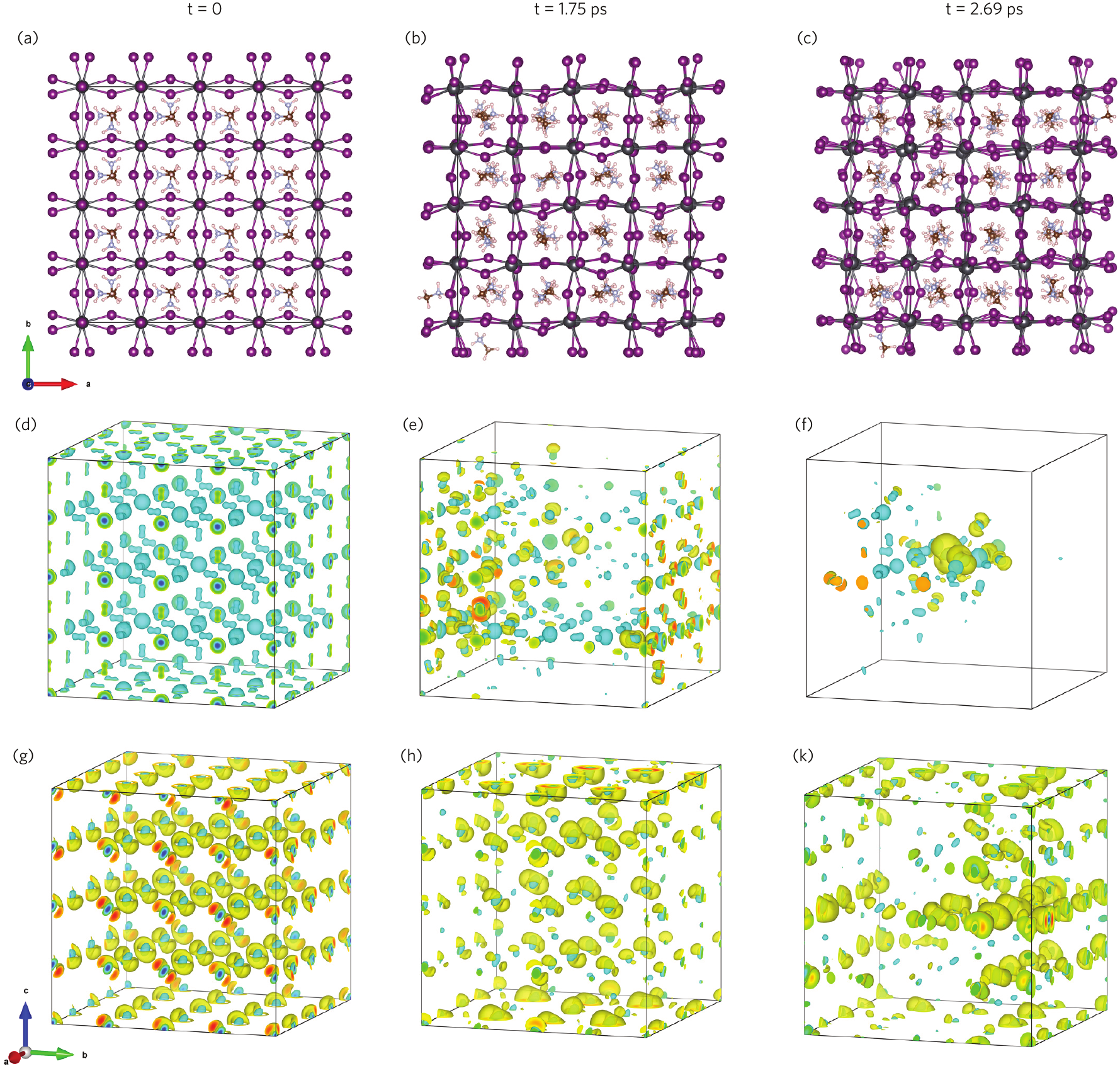}\\
	\caption{(a-c) $4\times4\times4$ tetragonal \ce{(CH3NH3)PbI3} supercell structure at different MD time steps and the corresponding spatial distribution of DFT orbitals $\psi_i(\bm{r})$ for (d-f) VBE and (g-k) CBE states. Isosurfaces correspond to the values of $\psi=\pm5\times10^{-4}$.}\label{fig-9}
\end{figure*}

For solar cells it is important to have a material with good \text{bipolar} transport properties. It is known that mobility decreases with $T$ due to electron-phonon scattering, which is also true for perovskites \cite{Milot_AFM_25_2015}. The question is whether the charge transport coefficients of electrons and holes are equally susceptible to thermal structural fluctuations. To address this question, we plot the wavefunction of electronic eigenstates at the VBE and CBE in Fig.~\ref{fig-9}. The structural fluctuations impact the spatial coherency of wavefunctions at the band edges. Electronic states at the VBE undergo the most significant changes that eventually leads to their spatial localization (Fig.~\ref{fig-9}f). Electronic states at the CBE seem to be more robust against thermal structural fluctuations, which is also consistent with their better ability to retain the Bloch character (Fig.~\ref{fig-7}).

\begin{figure}[t]
	\includegraphics{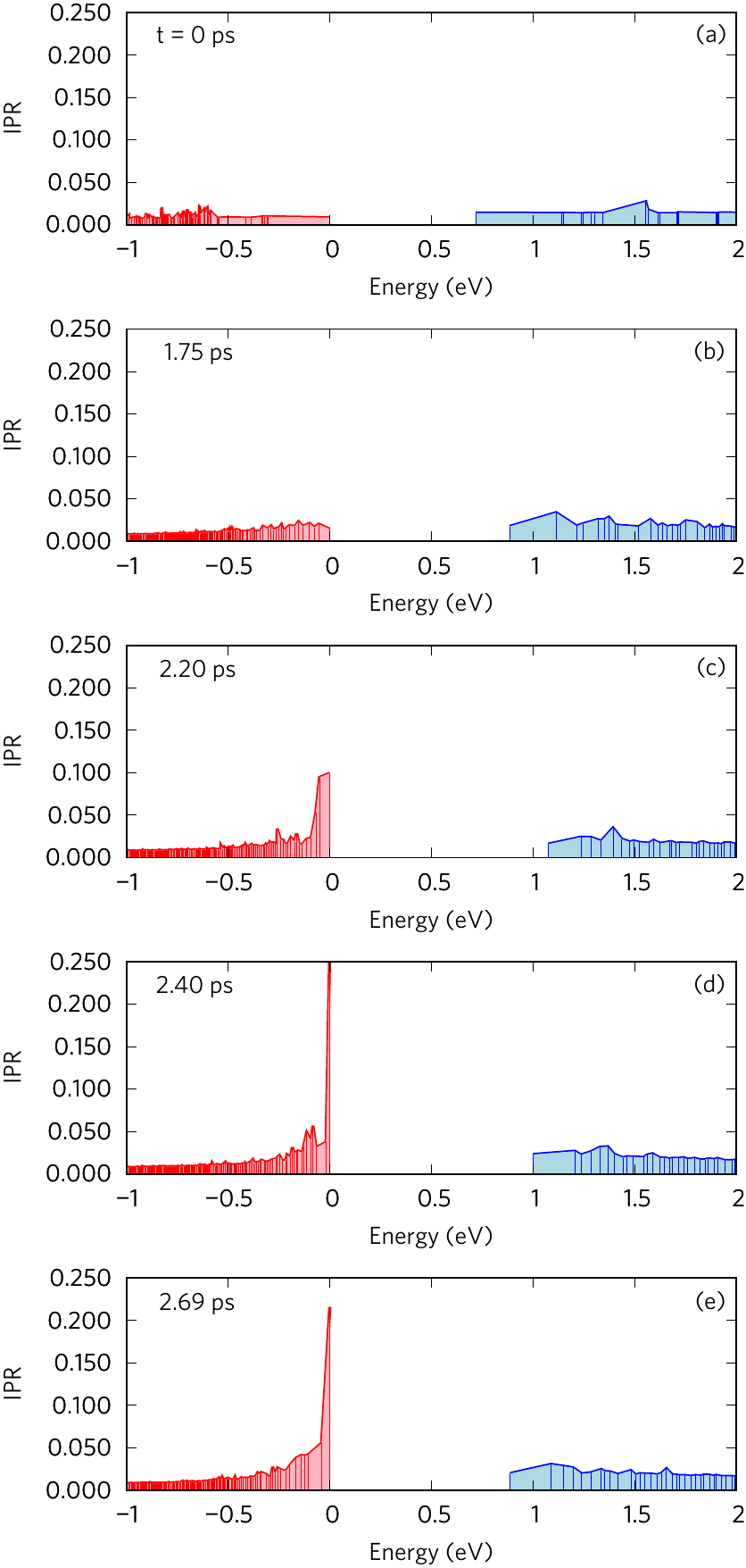}\\
	\caption{Spectrally-resolved inverse participation ratio for states in vicinity of the band gap of $4\times4\times4$ tetragonal \ce{(CH3NH3)PbI3} supercell taken at different MD time snapshots:  (a) $t=0$, (b) 1.75~ps, (c) 2.2~ps, (d) 2.4~ps and (e) 2.69~ps. A spatial localization of the DFT orbitals takes place at the valence band edge due to dynamic atomic displacements. The structural fluctuations open the band gap that continues to change dynamically.}\label{fig-10}
\end{figure}

To further explore possible localization of electronic eigenstates, we present IPR spectra in Fig.~\ref{fig-10} for structures at different MD time snapshots. IPR represents the inverse number of atoms that contribute to an eigenstate. Localized states exhibit higher IPR values. The initial structure (Fig.~\ref{fig-10}a) sets a baseline for the IPR spectrum in the case of a localization-free structure. The IPR spectrum shows a clear distinction between structures at 1.75 and 2.69~ps. A spike in the IPR spectrum at the VBE (Fig.~\ref{fig-10}e) confirms localization of these states. Additional spectra obtained at MD times of 2.2 and 2.4~ps (Fig.~\ref{fig-10}c,d) confirm that the localization in the VBE is a persistent feature, not a random occurrence. Thus we would expect the mobility of holes in \ce{(CH3NH3)PbI3} to be much more sensitive to thermal structural fluctuations (phonon scattering) than the mobility of electrons. This result is consistent with a steeper increase of the effective mass of holes with temperature predicted earlier \cite{Lu_APL_111_2017}.

By observing changes in the band gap in Fig.~\ref{fig-10}, we can conclude that the dynamic structural disorder opens up the band gap substantially as also noticed by \citet{Saidi_JPCL_7_2016} and \citet{McKechnie_PRB_98_2018}. After opening, the band gap continues to change dynamically with fluctuations of $\sim0.3$~eV that is consistent with literature  \cite{Saidi_JPCL_7_2016,Monserrat_arXiv_1711.06274}. It is interesting to note that the band gap renormalization of such a large magnitude due to the structural dynamic is inherent to wide band gap materials only. Furthermore, the renormalization typically leads to a \textit{decrease} of the band gap \cite{Cardona_RMP_77_2005} contrary to what is observed in perovskites where the gap widens.

\begin{figure*}[t]
	\includegraphics{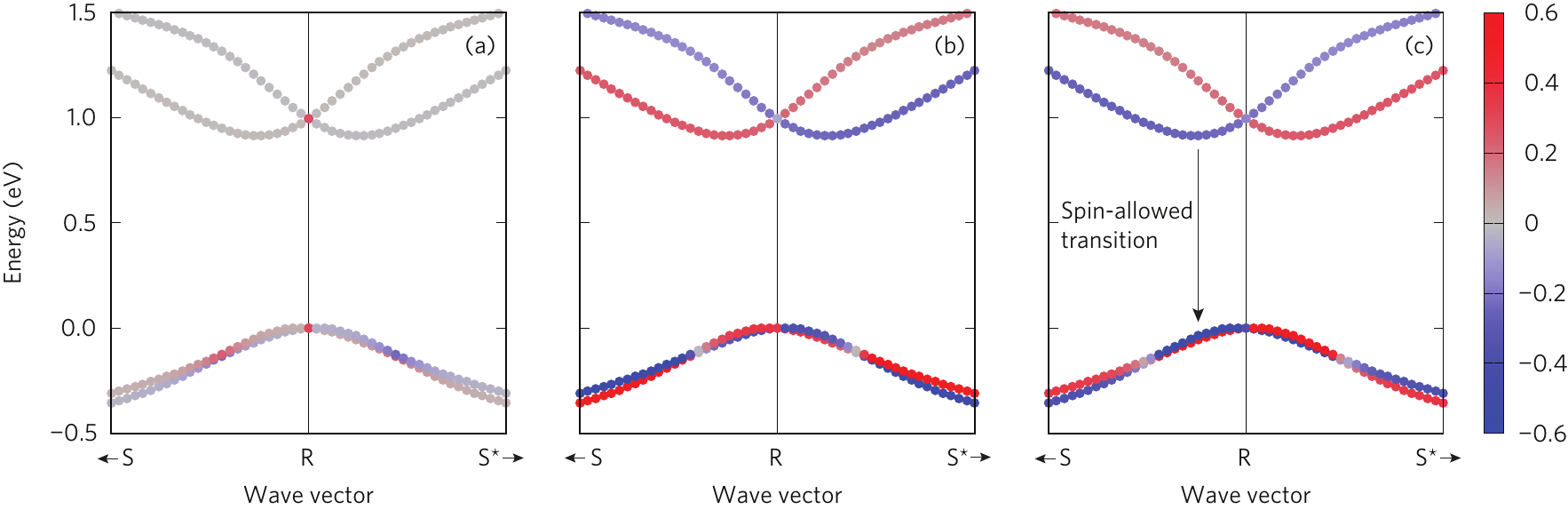}\\
	\caption{Spin texture (a) $\langle S_x \rangle$, (b) $\langle S_y \rangle$, and (c) $\langle S_z \rangle$ in a symmetrized unit cell of cubic \ce{(CH3NH3)PbI3} with Pb-atom displaced by 0.54~{\AA} along $[111]$ direction from the center of an octahedra formed by iodine atoms (Fig.~\ref{fig-5}a, CCDC deposition no.~1870792).}\label{fig-11}
\end{figure*}

Finally, we would like to comment on a spin texture at the band edges of perovskites. The long carrier lifetime is sometimes also attributed to spin-forbidden optical transitions rooted to the Rashba splitting \cite{Zheng_NL_15_2015}. Figure~\ref{fig-11} presents the band structure of cubic \ce{(CH3NH3)PbI3} ($\delta_\text{Pb}=0.54$~{\AA}, Fig.~\ref{fig-5}a) with three spin projections $\langle S_x \rangle$, $\langle S_y \rangle$, and $\langle S_z \rangle$. The Rashba valleys show a spin splitting with the symmetry $E(\bm{k},\bm{S})=E(-\bm{k},-\bm{S})$ leading to a spin helicity. Optical transitions between CBE and VBE are hindered, provided the initial and final states have an opposite spin helicity. However, this is not the case in Fig.~\ref{fig-11} where CBE and VBE have \textit{similar} spin projections. Using spin projections it is possible to determine spinor components ($\alpha$ and $\beta$, see Sec.~\ref{Sec:Method}) and the spin overlap at the band edges. Results presented in Table~\ref{Table:1} indicate an almost perfect spin overlap of 99\% between the two states representing CBE and the corresponding top of the valence band shown by arrow in Fig.~\ref{fig-11}c. The spin overlap remains high (greater than 90\%) in the tetragonal phase of \ce{(CH3NH3)PbI3} (Table~\ref{Table:1}). Examination of spin states in two dynamic structures ($t=0.42$ and $2.69$~ps, Table~\ref{Table:1}) reveals no changes in this trend. Thus, we reach a conclusion that the spin helicity poses no barriers for recombination of optical excitations in \ce{(CH3NH3)PbI3} contrary to Ref.~\onlinecite{Zheng_NL_15_2015}.

\begingroup
\squeezetable
\begin{table*}\label{Table:1}
\caption{Spin texture and spin overlap between the conduction band edge and the valence band in different structures of \ce{(CH3NH3)PbI3}.}\label{Table:1}
\begin{ruledtabular}
\begin{tabular}{lcccc}
Parameter & cubic & tetragonal & MD & MD \\
          & ($\delta_\text{Pb}=0.54$~{\AA}, Fig.~\ref{fig-5}a) &           &  ($t=0.42$~ps) & ($t=2.69$~ps) \\
\hline
Conduction band edge & & & & \\
\hspace{12pt}Wave vector $\bm{k}$ & $[0.44,1/2,1/2]$ & $[0.02,0,0]$ & $[0,1/2,0]$ & $[0,0,0.111]$ \\
\hspace{12pt}Unfolded $\bm{k}$ (Bloch character) & $-$ & $-$ & $[1/2,-0.375,1/2]$ (62\%) & $[1/2,1/2,0.472]$ (56\%) \\
\hspace{12pt}$\langle S_x \rangle$ &  $0.008$ & $-0.026$ & $-0.086$ & $0.154$ \\
\hspace{12pt}$\langle S_y \rangle$ &  $0.238$ & $0.192$ & $-0.001$ & $-0.028$ \\
\hspace{12pt}$\langle S_z \rangle$ &  $-0.245$ & $-0.089$ & $0.083$ & $-0.008$ \\
\hspace{12pt}Spin up $\alpha$ $\left(|\alpha|^2\right)$ & $-0.376+0i$ (14\%) & $-0.539+0i$ (29\%) & $-0.920+0i$ (85\%) & $0.690+0i$ (48\%) \\
\hspace{12pt}Spin down $\beta$ $\left(|\beta|^2\right)$ & $-0.030-0.926i$ (86\%) & $0.098-0.836i$ (71\%) & $0.392+0.006i$ (15\%) & $0.712-0.130i$ (52\%) \\
\hline
Valence band & & & & \\
\hspace{12pt}$\langle S_x \rangle$ &  $0.063$ & $-0.021$  & $-0.033$ & $0.228$ \\
\hspace{12pt}$\langle S_y \rangle$ &  $0.322$ & $0.367$  & $-0.035$ & $-0.043$ \\
\hspace{12pt}$\langle S_z \rangle$ &  $-0.524$ & $-0.452$  & $0.135$ & $-0.187$ \\
\hspace{12pt}Spin up $\alpha$ ($|\alpha|^2$) & $0.276+0i$ (8\%) & $0.335+0i$ (11\%) & $0.985+0i$ (97\%) & $0.432+0i$ (19\%) \\
\hspace{12pt}Spin down $\beta$ ($|\beta|^2$) & $0.185+0.943i$ (92\%) & $-0.054+0.941i$ (89\%) & $-0.118-0.123i$ (3\%) & $0.887-0.167i$ (81\%) \\
\hline
Spin overlap $|\langle \psi_\text{v} | \psi_\text{c} \rangle|^2$, \% & 99 & 95 & 91 & 90 \\
\end{tabular}
\end{ruledtabular}
\end{table*}
\endgroup

%
%
\section{Conclusions}\label{Sec:Conclusions}

The extended carrier lifetime in hybrid halide perovskites was attributed to a quasi-indirect band gap that arises due to Rashba splitting in both conduction and valence band edges. In this paper we calculated an effective relativistic band structure of \ce{(CH3NH3)PbI3} with the focus on the dispersion of electronic states near the band edges of \ce{(CH3NH3)PbI3}  affected by thermal structural fluctuations. The disorder is explicitly modeled via \textit{ab initio} molecular dynamics simulation performed for a large supercell. A semi-empirical scaling of the lattice parameters was used to achieve a finite-temperature structure of \ce{(CH3NH3)PbI3}.

Our preliminary studies involving static structures indicated that a tetragonal \ce{(CH3NH3)PbI3} with the \ce{PbI3} cage symmetry fixed to that obtained from x-ray diffraction studies shows no Rashba splitting even though the whole structure lacks the inversion center. This result is attributed to a centrosymmetric Pb-site symmetry. The fully relaxed tetragonal structure showed a weak Rashba splitting ($E_\text{R}=5$~meV), which occurs due to breaking a Pb-atom site inversion asymmetry. For the splitting energy to reach $k_\text{B} T\approx26$~meV at room temperature, the displacement should be of the order $0.3$~{\AA}, which is far above the static displacements of Pb-atoms in the tetragonal phase of \ce{(CH3NH3)PbI3}. Remarkably, lead-free perovskites structures (with Pb substituted by Sn while keeping geometry the same) show a comparable $E_\text{R}$ despite a drastic difference in the spin-orbit coupling constants.

The dynamic average displacement of Pb-atoms amounts to 0.37~{\AA} at room temperature with the amplitude reaching 0.8~{\AA}. The band structure of supercells taken at  different time snapshots during molecular dynamics simulation were unfolded to a primitive (pseudo-cubic) Brillouin zone. The most pronounced Rashba splitting of $E_\text{R}\approx120$~meV is observed when the dynamics of Pb-atoms is still coherent. The dynamic Rashba splitting diminishes down to 4~meV as the molecular dynamics progress to longer times (beyond 2~ps), which is assigned to randomization of Pb-displacement vectors and associated cancelation of the net effective magnetic field acting on electrons at the conduction band edge. At the same time, the valence band becomes disordered and loses the Bloch character that undermines the argument about an indirect band gap present in the dynamic structure of \ce{(CH3NH3)PbI3} at room temperature. Analysis of the spatial distribution of DFT orbitals at the valence band edge reveals their spatial localization in the dynamic structures as also confirmed by the inverse participation ratio spectra. Electronic states at the conduction band edge are more robust against thermal structural fluctuations, which is also consistent with their better ability to retain the Bloch character. Thus, the mobility of holes should be much more susceptible to phonon scattering than the mobility of electrons. The finite-temperature structural dynamics opens the band gap that continues to fluctuate with the amplitude of $\sim0.3$~eV at the sub-picosecond time scale. Analysis of spin projections and the spin overlap at the band edges rules out the spin helicity as a possible mechanism for a long lifetime of optical excitations in perovskite structures.

%
%
\begin{acknowledgments}
Authors are thankful to Dr.~Ray~LaPierre from the McMaster University for reading the manuscript. C.Z. and O.R. would like to acknowledge funding provided by the Natural Sciences and Engineering Research Council of Canada under the Discovery Grant Programs RGPIN-2015-04518. S.Y. would like to acknowledge funding provided by the Mitacs Globalink Research Internship program. DFT calculations were performed using a Compute Canada infrastructure supported by the Canada Foundation for Innovation under the John R. Evans Leaders Fund program.

C.Z. carried out MD simulations. S.Y. contributed the analysis of atomic displacements. O.R. designed the project, performed unfolded band structure calculations, and wrote the manuscript. All authors reviewed the manuscript before submitting for publication.
\end{acknowledgments}

%
%

\end{document}